\documentclass[10pt]{iopart2}

\usepackage{bm,mathrsfs,color}
\usepackage{graphicx}
\usepackage{amsmath}
\usepackage{color}

\begin{document}

\title[]{Negative transverse magnetoresistance due to negative off-diagonal mass in linear dispersion materials}

\author{Yudai Awashima$^1$ \& Yuki Fuseya$^{1, 2}$}

\address{
$^1$ Department of Engineering Science, University of Electro-Communications, Chofu, Tokyo 182-8585, Japan}
\address{$^2$ Institute for Advanced Science, University of Electro-Communications, Chofu, Tokyo 182-8585, Japan}
\ead{awashima@kookai.pc.uec.ac.jp}
\vspace{10pt}

\begin{abstract}
This study calculated the magnetoresistance (MR) in the Dirac electron system, Dresselhaus--Kip--Kittel (DKK) model, and nodal-line semimetals based on the semiclassical Boltzmann theory, with particular focus on the detailed energy dispersion structure.
The negative off-diagonal effective-mass was found to induce negative transverse MR owing to the energy dispersion effect.
The impact of the off-diagonal mass was more prominent in case of a linear energy dispersion.
Further, Dirac electron systems could realize negative MR even if the Fermi surface was perfectly spherical. The obtained negative MR in the DKK model may explain the long-standing mystery in p-type Si.

\end{abstract}

%
%
%
%
\ioptwocol

\section{Introduction}

Since its discovery by Thomson in 1857 \cite{Thomson_1857}, magnetoresistance (MR) has garnered attention from both scientific and practical use perspectives, such as mobility measurements and sensitive magnetic sensors \cite{Beer_1963,pippard_1989}.
The microscopic origin of orbital MR is the Lorentz force.
Therefore, the essential properties of orbital MR must be explainable within the semiclassical theory. 
However, the simple semiclassical theory provided in textbooks \cite{Kittel_1963,Ziman_1972} cannot comprehensively explain the properties of MR even in elementary systems.
For example, the simple semiclassical theory states that transverse MR (TMR) (current and magnetic field are perpendicular) is not observed in single carrier systems with parabolic dispersion.
However, experiments have confirmed that many single carrier systems exhibit a clear TMR \cite{Beer_1963,pippard_1989}.
Further, it states that longitudinal MR (LMR) (current and magnetic field are in parallel) should not be realized in every system owing to the absence of the Lorentz force in such a configuration.
However, many systems have been experimentally shown to exhibit a finite LMR, and in certain cases, the LMR has been found to be larger than TMR \cite{Allgaier_1958}.

Effective mass is considered a key aspect for understanding such unexpected properties. It is expressed as $\alpha_{ij}(=m_{ij}^{-1})=\hbar^{-2}\partial^2E(k)/\partial k_i\partial k_j$.
Conventional simple semiclassical theories assume a parabolic energy dispersion, where the effective mass is diagonal and constant ($k$-independent). However, the MR can exhibit unexpected properties in cases where the $k$-dependence or the off-diagonal components of the effective mass are involved.
For example, even for a single carrier system, the TMR has been observed owing to the $k$-dependent effective mass \cite{Awashima_2019,Collignon_2021}.
In addition, the LMR becomes finite when the effective mass contains off-diagonal elements.
Moreover, off-diagonal effective mass driven large LMR has been observed in multi carrier systems \cite{Mitani_2020}.

Recently, another anomalous MR, referred to as the negative MR has garnered renewed interest, particularly in Weyl semimetals from the perspectives of chiral anomaly and the Berry curvature\cite{Nielsen_1983,Burkov_2015,Son_2013}.
Following this trend, it was recently proposed that the negative off-diagonal mass can induce negative MR in Weyl semimetals, irrespective of the chiral anomaly or the Berry curvature \cite{Yamada_2022}.
However, whether this concept of negative off-diagonal mass is common or specific to Weyl semimetals is yet to be understood.

Thus, this study showed that the concept of negative MR owing to the negative off-diagonal mass is considerably common, particularly in systems with a linear dispersion. This was realized through the examination of three different systems: the Dirac electrons, Dresselhaus--Kip--Kittel (DKK) model, and nodal-line semimetals.
These particular systems were selected because their energy dispersions contain a root sign, which easily yields the negative off-diagonal mass.
For example, the energy dispersion of the isotropic Dirac electron system considered in this study is expressed as follows \cite{Wolff_1964,Fuseya_2015}: 
\begin{align}
\label{Dirac_dispersion}
     E^{\rm{Dirac}}_{\pm}=\pm\sqrt{\Delta^2+\Delta\frac{\hbar^2 k^2}{m^*}},
\end{align}
where $k^2=k_x^2+k_y^2+k_z^2$, $2\Delta$ is a band gap between the conduction and valence bands, $m^*$ is the effective-mass, and $\hbar$ is the Dirac constant.
The off-diagonal component of the inverse mass $\alpha_{xy}$ can be calculated as follows: 
\begin{equation}
\label{Dirac_axy_form}
 \alpha^{\rm{Dirac}}_{xy}=-\frac{\Delta^2 \hbar^2 k_xk_y}{m^{*2}\left(\Delta^2+ \Delta \hbar^2 k^2/m^*\right)^{3/2}}.
\end{equation}
As evident, $\alpha_{xy}^{\rm Dirac}$ has a negative sign, which is because of appearance of $k$-dependence in the root sign.
(Specifically, the positive or negative sign of the effective mass must be determined by excluding the factor $k_x k_y$, that is, by considering $\bar{\alpha}_{ij}=\alpha_{ij}k_ik_j/|k_i||k_j|$.)
Similarly, the DKK model and the nodal-line semimetals possess the negative off-diagonal mass; however, their $k$-dependences are different.
This study also discussed the influence of the linearity of the energy dispersion on the off-diagonal elements.

\section{Theory of magnetoresistance}
The MR was calculated considering a combination of the semiclassical Boltzmann theory and the well-known Chambers formula\cite{Shockley_1950, Chambers_1952, Chambers_1957, Ashcroft}.
Using the Chambers formula, the conductivity $\hat{\sigma}$ and resistivity $\hat{\rho}$ are expressed as
\begin{align}
\label{Chambers}
&\sigma_{ij}=e^2\langle v_i(\bm{k})\bar{v}_j(\bm{k})\rangle_F,\\
&\hat{\rho}=\hat{\sigma}^{-1}.
\end{align}
where $e$ ($>0$) is the elementary charge and $\langle\cdots\rangle_F=\int(d\bm{k}/4\pi^3)\cdots(-\partial f_0/\partial E)$ represents integration along the Fermi surface at low temperature ($f_0$ is the Fermi-Dirac distribution function at equilibrium).
Further, $\bm{v}(\bm{k})$ is the velocity of the electron obtained from the energy dispersion and is expressed as follows:
\begin{align}
\bm{v}(\bm{k})=\frac{1}{\hbar}\frac{\partial E(\bm{k})}{\partial \bm{k}}.
\end{align}
Moreover, $\bar{\bm{v}}(\bm{k})$ is the velocity averaged over time for electrons moving in a magnetic field.
\begin{align}
&\bm{\bar{v}}(\bm{k})=\int_{-\infty}^0dt\bm{v}(\bm{k}(t))\exp(t/\tau),\\
&\frac{d\bm{k}(t)}{dt}=-\frac{e}{\hbar}\bm{v}(\bm{k})\times\bm{B},
\end{align}
where $\bm{B}$ is the magnetic field and $\tau$ is the relaxation time. The Chambers formula does not assume any specific material in the derivation process and is thus applicable to any arbitrary energy dispersion.
However, the Chambers formula does not include effects derived from topological quantities such as the Berry curvature or chiral anomaly.
The following results were obtained based on the detailed structure of the energy dispersion without any topological effects.

\section{Dirac electron system}
\begin{figure}[t]
\begin{center}
\includegraphics[width=7cm]{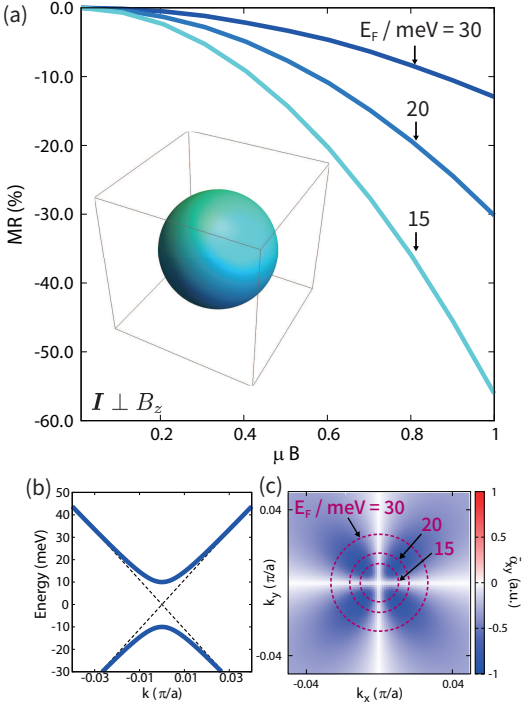}
\caption{(a) Calculated TMR in the Dirac electron system. Mass parameter $m^*=0.03$, band gap $\Delta=10 \ \rm{meV}$, $\mu$ is the mobility ($\rm{T^{-1}}$), and the product with the magnetic field $B$ ( $\rm{T}$), that is, $\mu B$ is a dimensionless quantity. The temperature is fixed at $1 \ \rm{K}$. The inset shows the spherical Fermi surface of an isotropic Dirac electron system. (b) Energy dispersion $E(k)$ of the Dirac electron system. The $k$ is normalized by $\pi/a$, where $a$ is the lattice constant. (c) Off-diagonal mass $\bar{\alpha}_{xy}$ mapped onto the $k_xk_y$ plane. The blue areas correspond to negative values. The dashed line represents the Fermi surface ($E_F=30,20,15 \ \rm{meV}$).}
\label{fig1}
\end{center}
\end{figure}

First, the single-carrier isotropic Dirac electron system was examined. Its energy dispersion is expressed as Eq. \eqref{Dirac_dispersion}. Owing to the Fermi surface of this model being perfectly spherical, no MR is expected upon the application of the parabolic dispersion approximation to this model \cite{Owada_2018}.
However, a large negative TMR was observed beyond the parabolic approximation.
Figure. \ref{fig1} (a) shows the TMR calculated using the Chambers formula for an isotropic Dirac electron system with a magnetic field applied in the $z$-direction, $\bm{B}=(0,0, B)$.
The calculations yielded a negative TMR, $-56\%$ at $\mu B= 1.0$ for $E_F=15 \ \rm{meV}$, where $\mu$ is the mobility. 
(maximum permissible value of the negative MR was $-100\%$).
Here, $\mu B=1.0$ corresponded to a relatively weak-field in Dirac electrons systems, where the effective mass was very small with a high mobility \cite{Allgaier_1958,Hartman_1969,Collaudin_2015,Zhu_2018}. For example, $\mu B=1.0$ corresponded to $B=0.01$ T when $\mu=100$ T$^{-1}$.
This finite TMR was caused by the $k$-dependence of the effective mass. 
Furthermore, the negative TMR originated from the negative off-diagonal mass.

The Chambers formula Eq. \eqref{Chambers} is not appropriate for determining the physical origin of the field-dependence in the resistivity because the contribution from $\bm{B}$ is hidden.
By contrast, to clarify the physical origin of the field-dependence, the Jones-Zener expansion is extremely useful \cite{Jones-Zener}.
Thus, the magneto-resistivity $\rho_{xx}$ was expanded up to the second order with respect to $B$ as follows:
\begin{align}
\label{expansion}
&\rho^{(2)}_{xx}=\tau^2\left[R_{xx}+R_{xy}\right] B^2,\\
\label{Rxx_a}
&R_{xx}=\Lambda_x^2\langle v_x^2(\alpha_{xx}\alpha_{yy}-\alpha_{xy}^2)\rangle_F\nonumber\\
&+\Lambda_x^2\langle v_x^2(v_x\gamma_{xyy})
+v_x^2(v_y\gamma_{xxy}+v_y\gamma_{xyx})
-v_x^2(v_x\gamma_{xxx})\rangle_F.\\
\label{Rxy}
&R_{xy}=-\Lambda_x^2\Lambda_y\langle(v_xv_y\alpha_{xy}-v_y^2\alpha_{yy})^2\rangle_F
\end{align}
where $\Lambda_{\mu}=1/\langle v_{\mu}\rangle_F^2$ and $\gamma_{ijk}=\hbar^{-3}\partial^3E(k)/\partial k_i\partial k_j\partial k_k$.
$R_{xx}$ originates from $\sigma_{xx}(B)$, and $R_{xy}$ originates from $\sigma_{xy}^2(B)$.
Consequently, it is evident that the coefficient of $B^2$ is mostly obtained in terms of the (invers) effective mass $\alpha_{ij}$.
Therefore, the intuitive understandings for the anomalous MR can be obtained upon an in-depth investigation of the properties of the effective-mass.

The key to the negative TMR are $R_{xy}$ and $\alpha_{xy}$, where the former negatively contributes to $\rho_{xx}$ owing to the negative coefficient in Eq. \eqref{Rxy}. In the ordinary case of $\bar{\alpha}_{xy} >0$, the negative contribution from $R_{xy}$ is small because the $\alpha_{yy}$-term offsets the $\alpha_{xy}$-term.
However, when $\bar{\alpha}_{xy}<0$, both the $\alpha_{xy}$ and $\alpha_{yy}$ cooperatively boost the negative contribution of $R_{xy}$.
Consequently, the negative off-diagonal mass can realize the negative TMR.

The Dirac electron system is a typical example for such a negative off-diagonal mass induced negative TMR.
Figure. \ref{fig1} (c) shows a plot $\bar{\alpha}_{xy}$ as a function of $k_x$ and $k_y$. 
The majority of $\bar{\alpha}_{xy}$ was negative. Further, $\bar{\alpha}_{xy}$ was vanishingly small when $k_{x, y}=0$, although it was not positive.
The widespread negative region in $\bar{\alpha}_{xy}$ causes the significantly negative TMR in the Dirac electron system.

Moreover, the TMR of the Dirac electron system becomes zero in the region $\Delta\simeq E_F$, which can be approximated as the parabolic dispersion. 
The energy dispersion Eq. \eqref{Dirac_dispersion} can be approximated as 
\begin{align}
 E^{\rm{Dirac}}_{+}\simeq\Delta+\frac{\hbar^2}{2m}\left(k_x^2+k_y^2+k_z^2\right),
\end{align} 
for $\Delta\simeq E_F$.
The off-diagonal mass becomes zero in case of this parabolic dispersion, and the $k$-dependence of the diagonal element vanishes.
Consequently, the negative TMR should be vanishingly small when $E_F$ approaches $\Delta$.

\section{DKK model}
\begin{figure}[t]
\begin{center}
\includegraphics[width=7cm]{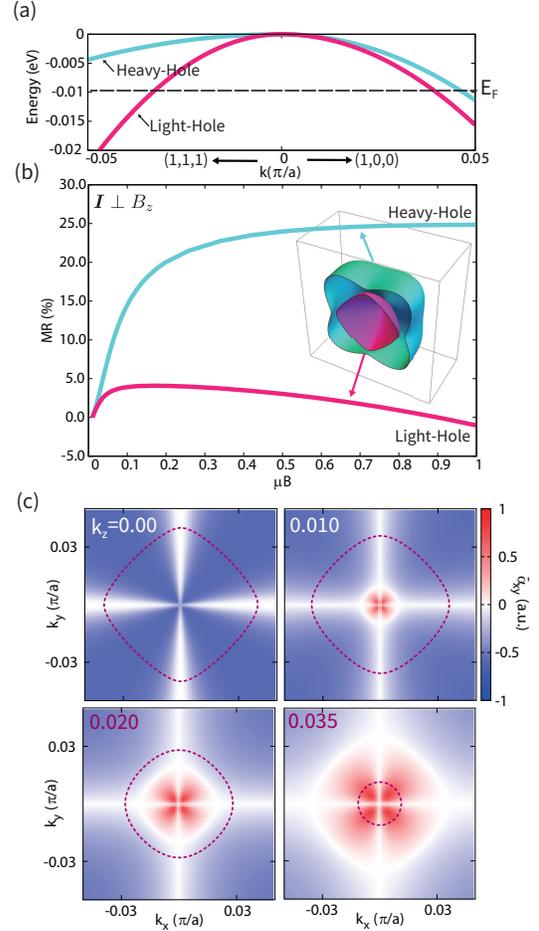}
\caption{(a) Energy dispersion $E(k)$ of the DKK model.(b) Calculated TMR of the DKK model. Temperature is fixed at $1 \ \rm{K}$. The diagram of the Fermi surface shows a cross-sectional view of a Fermi surface composed of outer heavy and inner light hole, cut by a plane with $k_y=0$. (c) Plot of the $\bar{\alpha}_{xy}$ of light hole mapped onto the $k_xk_y$ plane. The dashed line is the Fermi surface of the light hole at $E_F = -0.01 \ \rm{eV}$ corresponding to the value of $k_z$.}
\label{fig2}
\end{center}
\end{figure}

The DKK model is an effective model for investigating the valence band around the $\Gamma$ point of Si and Ge.
The energy dispersion is expressed as\cite{Dresselhaus_1954}
\begin{align}
E^{\rm{DKK}}_{\pm}=Ak^2\pm\sqrt{B^2k^4+C^2(k_x^2k_y^2+k_y^2k_z^2+k_z^2k_x^2)}.
\end{align}
It is expressed in a form that included the root sign similar to that in case of the Dirac electron system; that is, the negative TMR because of the negative off-diagonal mass is expected.
In Si, the parameters were $A=-4.29$, $B=0.68$, and $C=4.87$ in units of $\hbar^2/2m$\cite{Kittel_2004}.
Further, positive and negative signs corresponded to heavy and light hole, respectively \cite{Beer_1963}.
Figure. \ref{fig2} (a) shows the energy dispersion of DKK model.
The single carrier MR was calculated for each heavy and light hole shown in Fig. \ref{fig2} (b).
The MR was always positive for heavy hole and saturated at $25\%$.
However, for light hole, the MR increased up to $\mu B=0.2$ and exhibited saturation signs at approximately $5\%$, after which it began to decrease and became negative at approximately $\mu B=1.0$.

In the presence of both positive and negative off-diagonal masses, the behaviour of the MR is governed by the one dominant in the entire Fermi surface.
The effective mass $\alpha_{xy}$ in the DKK model is expressed as
\begin{align}
\label{DKK_axy}
&\alpha^{\rm{DKK}}_{xy}=A_{x}A_{y}/(4\hbar^2(D^{\rm{DKK}})^3)\\
&\qquad\pm k_xk_y(8B^2+4C^2)/(2\hbar^2 D^{\rm{DKK}}),\nonumber\\
&A_{x}=4B^2k^2k_x+C^2k_x(2k_y^2+2k_z^2),\nonumber\\\
&A_{y}=4B^2k^2k_y+C^2k_y(2k_x^2+2k_z^2),\nonumber\\
&D^{\rm{DKK}}=(B^2k^2+C^2(k_x^2k_y^2+k_y^2k_z^2+k_z^2k_x^2))^{1/2}.\nonumber
\end{align}
Figure. \ref{fig2} (c) shows a plot of $\bar{\alpha}_{xy}$ of light hole against several $k_z$.
Negative off-diagonal element was observed over a wide range; however, positive values appeared for certain values of $k_z$.
The entire Fermi surface, including all $k_z$, was dominated by negatives; however, positive values weakened the negative off-diagonal element.
The opposite was true for heavy hole, which exhibited a negative off-diagonal element in certain parts; however, a positive off-diagonal element was observed in the entire Fermi surface.
Thus, a heavy hole with a positive dominant off-diagonal element results in a positive MR. By contrast, a light hole with a dominant negative element yields a negative MR.

\section{Nodal-line semimetals}
\begin{figure}[t]
\begin{center}
\includegraphics[width=7cm]{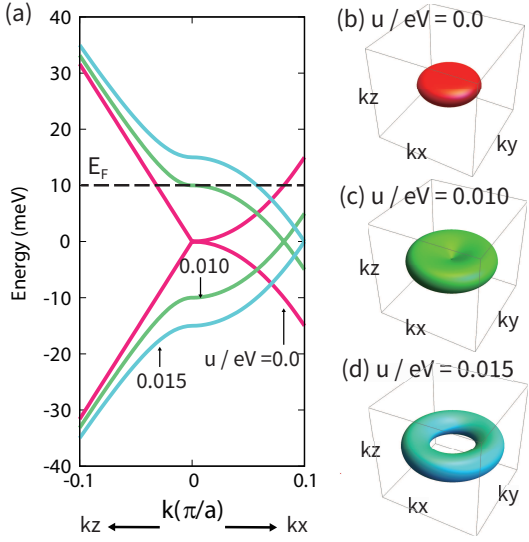}
\caption{(a) Energy dispersion of the effective model in nodal-line semimetals. $\lambda=0.32 \ \rm{eV.\AA}$. (b--d) Fermi surfaces at each parameter when fermi energy $E_F=0.01 \ \rm{eV}$.}
\label{fig3}
\end{center}
\end{figure}

In nodal-line semimetals, the band gap is closed along the nodal-line in $\bm{k}$ space \cite{Fujioka_2019, Weng_2015, Neupane_2016, Chen_2017}.
The following Hamiltonian was proposed for nodal-line semimetals as an effective model obtained using the $\bm{k}\cdot \bm{p}$ theory\cite{Li_2018, Bian_2016-1, Bian_2016-2}.
\begin{align}
 H^{\rm{nodal}}=\left\{\left[\frac{\hbar^2}{2m}(k_x^2+k_y^2)-u\right]\tau_3+\lambda k_z\tau_1\right\}\otimes\sigma_0.
\end{align}
where $\hat{\sigma}$ and $\hat{\tau}$ are the Pauli matrices in the spin and orbital spaces, respectively, and $m$, $u$, and $\lambda$ are the model parameters.
The eigenvalue is expressed as
\begin{align}
 E^{\rm{nodal}}_{\pm}=\pm\sqrt{\left(\frac{\hbar^2}{2m}(k_x^2+k_y^2)-u\right)^2+\lambda^2k_z^2}.
\end{align}
The energy dispersion is shown in Fig. \ref{fig3} (a).
Parameter $u$ exhibits an energy dimension and corresponds to the band gap size at the $\Gamma$ point.
For $u<E_F$, the Fermi surface was a drumhead Fermi surface [Fig. \ref{fig3} (c)], and for $u>E_F$, it was a torus shaped Fermi surface [Fig. \ref{fig3} (d)].

\begin{figure}[t]
\begin{center}
\includegraphics[width=7cm]{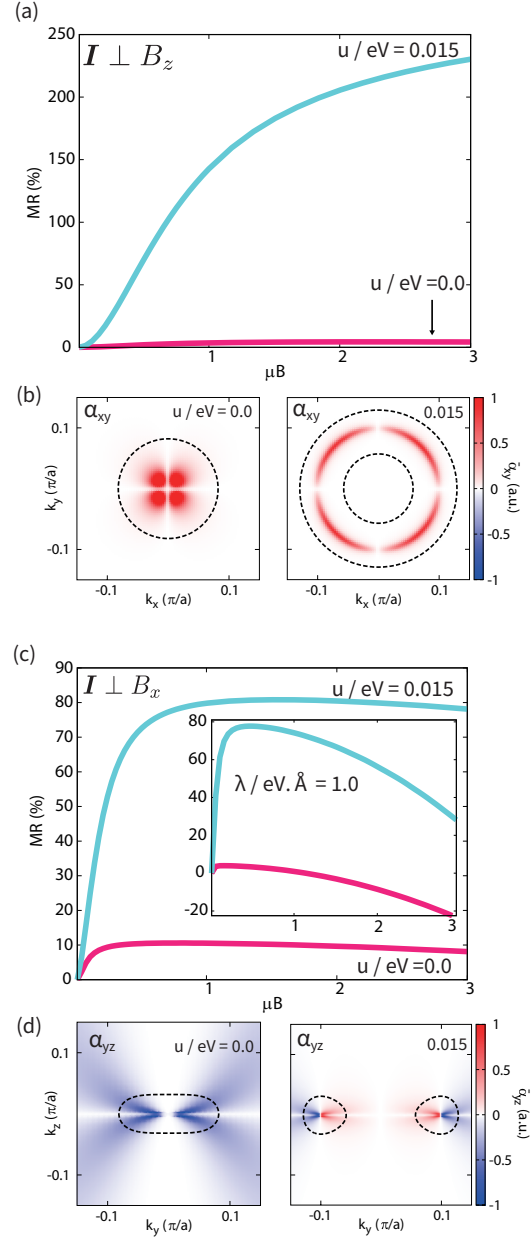}
\caption{(a)TMR($\bm{B} \parallel z$) of nodal-line semimetals at $\lambda=0.32 \ \rm{eV.\AA}$. Temperature is fixed at $1 \ \rm{K}$. (b)Plot of the off-diagonal mass $\bar{\alpha}_{xy}$ in the $k_xk_y$ plane. The dashed line is the Fermi surface at $E_F=0.01 \ \rm{eV}$. (c)TMR($\bm{B} \parallel x$) of nodal-line semimetals at $\lambda=0.32 \ \rm{eV.\AA}$. The inset shows TMR at $\lambda=1.0 \ \rm{eV.\AA}$. Temperature is fixed at $1 \ \rm{K}$. (d)Plot of the off-diagonal mass $\bar{\alpha}_{yz}$ in the $k_yk_z$ plane.}
\label{fig4}
\end{center}
\end{figure}

As MR results from Lorentz forces, the off-diagonal mass perpendicular to the applied magnetic field is essential.
For example, $\alpha_{xy}$ for $\bm{B}=(0, 0, B)$ and $\alpha_{yz}$ for $\bm{B}=(B, 0, 0)$ are important factors that affect the MR.
The model used in this study was anisotropic in the $k_z$ and $k_x,k_y$ directions.
Therefore, the effective-masses $\alpha^{\rm{nodal}}_{xy}$ and $\alpha^{\rm{nodal}}_{yz}$ differed significantly from each other.
The off-diagonal elements $\alpha^{\rm{nodal}}_{xy}$ and $\alpha^{\rm{nodal}}_{yz}$ of $E^{\rm{nodal}}_{+}$ are
\begin{align}
&\alpha^{\rm{nodal}}_{xy}=\lambda^2\frac{4\hbar^2k_xk_yk_z^2}{(2m)^2D^{\rm{nodal}}},\\
\label{li-alpha_yz}
&\alpha^{\rm{nodal}}_{yz}=-\lambda^2\frac{k_yk_z\left[ \hbar^2(k_x^2+k_y^2)/2m-u\right]}{mD^{\rm{nodal}}},\\
&D^{\rm{nodal}}=\left\{\left[\frac{\hbar^2}{2m}(k_x^2+k_y^2)-u\right]^2+\lambda^2 k_z^2\right\}^{3/2}.\nonumber
\end{align}

In this effective model of the nodal-line semimetals, the behaviour of the TMR differed significantly depending on the direction of the magnetic field.
Herein, the TMR characteristics were calculated for two magnetic field directions: $\bm{B}\parallel z$ and $\bm{B} \parallel x$.
The TMR for $\bm{B}\parallel z$ was positive MR for both $u=0.0,\ 0.015\ \rm{eV}$, particularly for $u=0.015 \ \rm{eV}$, reaching $230\%$ at $\mu B=3.0$ [Fig. \ref{fig4} (a)].
Further, $\bar{\alpha}^{\rm{nodal}}_{xy}$ was always positive for the entire region, as shown in Fig. \ref{fig4} (b).
By contrast, the TMR for $\bm{B}\parallel x$ decreased after reaching $80\%$ for $u=0.015 \ \rm{eV}$ and $10\%$ for $u=0.0$ eV [Fig. \ref{fig4} (c)].
The inset in Fig. \ref{fig4} (c) shows the calculation results when $\lambda=1 \ \rm{eV.\AA}$.
Compared to the case with $\lambda=0.32 \ \rm{eV.\AA}$, the decreasing trend of MR was stronger and a negative MR was obtained at approximately $\mu B=1.0$.
In contrast to the case of $\bm{B}\parallel z$, $\bar{\alpha}^{\rm{nodal}}_{yz}$ exhibited a negative region, particularly for $u=0.0$ [Fig. \ref{fig4} (d)].
This negative off-diagonal mass was the source of the decreasing TMR for $\bm{B} \parallel x$.

\section{Discussion}
\subsection{Linearity}
The impact of the negative off-diagonal mass on the MR is observed to be more significant for systems with linear dispersion.
This study demonstrated the reason why the linear dispersion induced negative MR compared with the quadratic dispersion, through simple examples. 
Suppose that the linear and quadratic energy dispersions $E^L$ and $E^Q$, respectively, are expressed as
\begin{align}
&E^{L}=\sqrt{\Delta^2+\Delta k^2},\\
&E^{Q}=\sqrt{\Delta^2+k^4}.
\end{align}
(Here, $\Delta \hbar^2/m=1$ was set for simplicity.)
Each off-diagonal mass is expressed as follows:
\begin{align}
&\alpha_{xy}^L=-\frac{\Delta^2 k_xk_y}{(\Delta^2+\Delta k^2)^{3/2}} 
,\\
&\alpha_{xy}^Q=-\frac{4k^2k_xk_y}{(\Delta^2+k^4)^{3/2}}+\frac{4k_xk_y}{(\Delta^2+k^4)^{1/2}} 
.
\end{align}
With linear dispersion, only the negative term appeared. Whereas, with quadratic dispersion, both negative and positive terms were found to coexist, thus offsetting each other.
Therefore, the magnitude of the negative MR for the Dirac electrons (linear) was larger than that for the DKK model (quadratic).
Further, nodal line semimetals exhibited linear dispersion for one direction ($k_z$) and quadratic dispersion for two directions ($k_x, k_y$). Consequently, the nodal line semimetals exhibited behaviour in-between that of the Dirac electrons and the DKK model.

\subsection{Comparison with experiments}
A primary finding of the study is that the Dirac electrons exhibited significant negative TMR.
However, typical Dirac electron systems, such as Bi and PbTe, have not been experimentally observed to exhibit negative TMR \cite{Hartman_1969,Collaudin_2015,Zhu_2018,Allgaier_1958}.
The source of the discrepancy between theory and experiment is the multi-carrier nature.
Bi and PbTe have three and four Dirac valley at the $L$ points in their Brillouin zone, respectively\cite{Fuseya_2015,Akiba_2018}.
To examine the manner in which the results of this study change in multi-carrier systems, the calculations performed on the Dirac electrons were extended to the four-valley system, which is an effective model for PbTe \cite{Mitani_2020}.

\begin{figure}[t]
\begin{center}
\includegraphics[width=7cm]{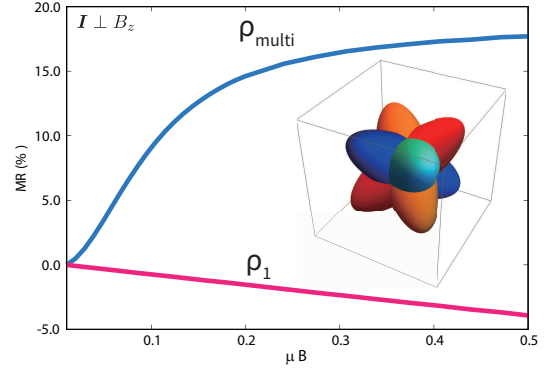}
\caption{Calculated TMR of the four-Dirac-valley model. $m^* = 0.1$, $\gamma=0.39$. Temperature is $1 \ \rm{K}$. $\rho_{\rm{multi}}$ is the TMR including four-Dirac-valley and $\rho_1$ is the TMR of a single carrier at $E^{\rm{Multi}}_1$. Inset shows a Fermi surface of the four-Dirac-valley model.}
\label{fig5}
\end{center}
\end{figure}

Four valleys of Dirac electrons were considered, where the long-axis of each Fermi surface was oriented along the four body-diagonal directions of the cubic (cf. the inset of Fig. \ref{fig5}). 
The energy dispersion of the four-Dirac-valley model is expressed as
\begin{align}
\label{4-valley_dispersion}
&E_{i}^{\rm{multi}}=\sqrt{\Delta^2+\Delta\frac{\hbar^2}{m^*}\bm{k}\cdot\hat{A}_i\cdot \bm{k}}\\
&\hat{A}_1=
\begin{pmatrix}
1&-\beta&-\beta\\
-\beta&1&-\beta\\
-\beta&-\beta&1
\end{pmatrix} ,\
\hat{A}_2=
\begin{pmatrix}
1&-\beta&\beta\\
-\beta&1&\beta\\
\beta&\beta&1
\end{pmatrix} ,\nonumber\\
&\hat{A}_3=
\begin{pmatrix}
1&\beta&\beta\\
\beta&1&-\beta\\
\beta&-\beta&1
\end{pmatrix} ,\
\hat{A}_4=
\begin{pmatrix}
1&\beta&-\beta\\
\beta&1&\beta\\
-\beta&\beta&1
\end{pmatrix}.
\end{align}
where $\beta$ in $\hat{A}_i$ is an anisotropic parameter and acquires the value of $0\leq \beta\leq 1/2$.
Further, it is fully isotropic at $\beta=0$ and fully anisotropic (i.e., a cylindrical Fermi surface) at $\beta=1/2$.
The total MR in the four-Dirac-valley model is obtained as
$\hat{\rho}_{\rm{multi}}=\left(\sum_i^4\hat{\sigma}_i\right)^{-1}$.

Figure. \ref{fig5} shows the TMR calculated in the four-valley model. 
When only one Dirac carrier was considered, its MR exhibited a negative TMR.
However, when simultaneously considering four Dirac carriers, the total MR became a positive TMR.

This transformation from negative to positive TMR can be understood on the basis of the off-diagonal mass.
In the multi-ellipsoidal model, the off-diagonal mass becomes finite and positive when the axis of ellipsoid is tilted from the orientation of the electric current \cite{Mitani_2020}.
This positive off-diagonal mass owing to the multi-carrier contributes positively to the TMR.
Consequently, in the multi-Dirac system, the positive off-diagonal mass owing to the multi-carrier nature suppresses the negative off-diagonal mass because of the linear dispersion, thus resulting in the transformation from negative to positive TMR.
Therefore, the negative TMR of Dirac electrons should be observed only in the single-carrier Dirac electron systems, where the Dirac electrons are located at the $\Gamma$ point.

The negative TMR has been already observed in p-type Si 60 years ago \cite{Pollak_1963,Long_1958,Roth_1963}. However, there exists no reasonable explanation for p-type Si exhibiting the negative TMR. Thus, the mechanism of the negative TMR remains a longstanding mystery.
The results for the DKK model may entail a reasonable explanation for this mystery. The negative off-diagonal mass of light hole yields negative TMR, and its negative contribution dominates the total TMR because the mobility of the light hole is much larger than that of the heavy hole.

Certain nodal-line semimetals exhibit negative LMR and positive TMR \cite{Li_2018_NMR,An_2019,Chen_2020}.
However, to the best of our knowledge, no experimental study exhibiting negative TMR has been reported.
A reason of this discrepancy between experiments and our theory can be attributed to the sensitivity to the direction of the magnetic field.
As shown in Sec. 5, the properties of the TMR change drastically according to the orientation of the magnetic field. 
Therefore, the negative TMR may be observed via the tuning of the magnetic field direction.
Another possible reason for the discrepancy is the multi-carrier nature of actual nodal-line semimetals.
The nodal-line semimetals generally have a very complex band structure, and they frequently contain other carriers in addition to the nodal-line carries\cite{Pan_2018,Bannies_2021}. 
This multi-carrier nature induces the positive TMR as we have shown for the multi-Dirac model.
To comprehensively explain the MR experiments in the actual nodal-line semimetals, their complex band structure must be considered based on the Chambers formula, which will be a future problem.
Moreover, if an ideal nodal-line material is discovered, the positive to negative TMR transformation can be observed by controlling the field orientation.

\section{Summary}
This study showed that the negative TMR can be realized in Dirac electrons, the DKK model, and the nodal-line semimetals based on the semi-classical Boltzmann theory with the Chambers formula.
It was emphasized that the negative TMR was induced by the negative off-diagonal mass, considering the detailed band structure and $k$-dependence of the effective-mass.

The effect of the negative off-diagonal mass was enhanced in the case of linear energy dispersion.
Further, the proposed negative TMR mechanism easily explained the negative TMR in p-type Si, which has been a long-standing mystery.
In addition, in Si and many other systems, the mysterious negative MR properties were found to be explainable by focusing on the negative off-diagonal mass, which could be easily obtained from the density functional theory.
Further research is required to clarify the relationship between effective-mass and MR in many other systems and deepen our understanding of negative MR.
Moreover, the proposed concept of negative off-diagonal mass is general in nature, and thus, future studies will focus on providing a new route to the negative MR in addition to the chiral anomaly and the Berry curvature.

\section*{Acknowledgements}
We would like to thank M. Tokunaga, A. Yamada, K. Behnia, B. Fauqu\'e, and Z. Zhu for the fruitful discussions and helpful comments.
This work was supported by JSPS KAKENHI grants 19H01850 and 22K18318.
\section*{Reference}
\bibliographystyle{iopart-num}
\bibliography{biblio}

\end{document}